\documentclass[12pt]{article}
\usepackage{amsfonts}
\usepackage{amssymb}
\usepackage{graphics}
\begin{document}
{\LARGE \centerline{Integral-free Wigner functions}}

\phantom{aaa}

\phantom{aaa}

{\large \centerline{A. Te\u{g}men}}%
\centerline{Physics Department, Ankara University, 06100 Ankara,
TURKEY}%
\centerline{{\it tegmen@science.ankara.edu.tr}}

\begin{abstract}
Wigner phase space quasi-probability distribution function is a
Fourier transform related to a given quantum mechanical wave
function. It is shown that for the wave functions of type $\psi
(q)=e^{-aq^2}\phi (q)$, the Wigner function can be defined in
terms of differential operators acting on a given function,
independently from the integral formula which appears in the
standard definition. Gaussian wave packet, harmonic and singular
oscillators are given as the examples.
\end{abstract}
\noindent\rule{7in}{0.01in}
\section{Introduction}
In a study of quantum corrections to classical statistical
mechanics, Wigner constructed the function
\begin{eqnarray}\label{wigner}
W(q,p)=\frac{1}{2\pi}\int_{-\infty}^{\infty}\;{\psi}^{*}(q-\frac{\hbar}{2}y)
\;{\psi}(q+\frac{\hbar}{2}y)\;e^{-iyp}\;dy.
\end{eqnarray}
that now called by his name \cite{ref:Wigner}. The goal was to
replace the quantum mechanical wave function $\psi(q)$ with a
probability distribution in phase space that providing a
particlelike description of the underlying wave propagation. Hence
the Wigner function (WF) exhibits a representation of quantum
states in phase space. Although in quantum mechanics due to the
uncertainty principle, a certain point in phase space does not
make sense, projection of the WF onto the $q$ and $p$ planes
generates the relevant marginal distributions associated with that
quantum state, where possible negative values correspond to
non-classical situations \cite{ref:Hillery}.

WF has been much studied theoretically and experimentally since
its introduction, not only in the context of quantum mechanics,
but also in various branches of physics and related sciences.

The WF given by (\ref{wigner}) can be obtained easily if the
integral allow to get an answer analytically. In most cases,
despite the existence of wave function $\psi(q)$, the integral
becomes quite cumbersome and sometimes it is impossible to handle
it.

The aim of the present paper is to present an alternative
definition of the WF for the wave functions of a certain type,
that is a derivational approach rather than the original integral
based one. This new definition gives, at least partly,
calculational tricks to overcome the difficulties mentioned above.

The approach considers a class of wave functions given in the form
\begin{eqnarray}\label{choice}
\psi(q)=e^{-a\,q^2}\phi(q)\qquad \qquad (a\;\neq 0),
\end{eqnarray}
where $a$ is a positive constant, and is based on the series
expansion of $\phi (q)$ in term of Hermite polynomials. In order
to establish the convergence of the series we need the following
theorem \cite{ref:Lebedev}:
\\If $\phi(q)$ is a piecewise smooth real function in every finite
subinterval $[-b,b]$ and if the integral
\begin{eqnarray}\label{cond}
\int_{-\infty}^{\infty} e^{-q^2}\phi^2(q) dq
\end{eqnarray}
is finite, then the series
\begin{eqnarray}\label{power}
\phi(q)=\sum_{n=0}^{\infty}C_n H_n(q), \qquad -\infty < q <
\infty,
\end{eqnarray}
where the coefficients $C_n$ can be computed by the orthogonality
properties of the Hermite polynomials, converges to $\phi (q)$ at
every continuity point of $\phi (q)$ . It is also assumed that in
the case of complex $\phi (q)$ the consequent of the theorem is
still valid {\it i.e.}, $\phi (q)$ admits a uniformly convergent
expansion on the basis of Hermite polynomials.

With these assumptions it is possible to replace the integral by
the series and finally to define the WF as the following
\begin{eqnarray}\label{main}
W(q,p)=\frac{1}{\hbar\sqrt{2\pi
a}}\;e^{-2aq^2}\phi^{*}\left(q-\frac{i\hbar}{2}\partial_p\right)
\;\phi\left(q+\frac{i\hbar}{2}\partial_p\right)\;
e^{-p^2/2a\hbar^2},
\end{eqnarray}
which is the main result of this paper. Note that the assumptions
ensure that $\phi (q)$ is smooth enough to be evaluated over the
differential operators. It is easy to show that the operators
$\phi$ and $\phi^*$ in (\ref{main}) are commutative and their
product is real. A useful aspect of this definition is that if
$\phi(q)$ contains a finite polynomial alone that may be an
orthogonal polynomial such as Hermite or (associated) Laguerre
polynomial, then obtaining the WFs for the individual states
requires only simple differentiations without needing integral
tables or computers.
\section{Derivation}
With the choice of (\ref{choice}) the WF (\ref{wigner}) amounts to
\begin{eqnarray}\label{wigner2}
W(q,p)=\frac{1}{2\pi}\,e^{-2aq^2}\int_{-\infty}^{\infty}dy\;{\phi}^{*}(q-\frac{\hbar}{2}y)
\;{\phi}(q+\frac{\hbar}{2}y)\;e^{-\frac{\hbar^2}{2}ay^2-iyp}\;dy.
\end{eqnarray}
For our purpose, the Hermite polynomials takes the form
\begin{eqnarray}\label{hermite}
H_n(q\pm \frac{\hbar}{2}y)=\sum_{r=0}^{[n/2]}
\frac{(-1)^r\,n!}{r!\,(n-2r)!}2^{n-2r}(q\pm
\frac{\hbar}{2}y)^{n-2r}.
\end{eqnarray}
By the substitution of (\ref{power}) and (\ref{hermite}) into
(\ref{wigner2}) we get
\begin{eqnarray}\label{int}
W(q,p)&=&\frac{1}{2\pi}\;e^{-2aq^2}\sum_{n,m=0}^{\infty}C_n^*\;C_m\;
\sum_{r=0}^{[n/2]}\frac{(-1)^r\,n!}{r!\,(n-2r)!}2^{n-2r}
\sum_{s=0}^{[m/2]}\frac{(-1)^s\,m!}{s!\,(m-2s)!}2^{m-2s}\nonumber\\
& \times & \int_{-\infty}^\infty (q-\frac{\hbar}{2}y)^{n-2r}
(q+\frac{\hbar}{2}y)^{m-2s}\;e^{-\frac{\hbar^2}{2}ay^2-iyp}\;dy.
\end{eqnarray}
If one uses the Binomial expansion for the terms in the integral
in (\ref{int}), the WF takes the form
\begin{eqnarray}\label{xp}
W(q,p)&=&\frac{1}{2\pi}\;e^{-2aq^2}\sum_{n,m=0}^{\infty}C_n^*\;C_m\;
\sum_{r=0}^{[n/2]}\frac{(-1)^r\,n!}{r!\,(n-2r)!}2^{n-2r}
\sum_{s=0}^{[m/2]}\frac{(-1)^s\,m!}{s!\,(m-2s)!}2^{m-2s}\nonumber\\
& \times & \sum_{\alpha=0}^{n-2r}\left(
\begin{array}{c} n-2r \\
\alpha
\end{array}
\right)q^{n-2r-\alpha}(-\frac{\hbar}{2})^\alpha \;
\sum_{\beta=0}^{m-2s}\left(
\begin{array}{c} m-2s \\
\beta
\end{array}
\right)q^{m-2s-\beta}(\frac{\hbar}{2})^\beta \nonumber\\
& \times & \underbrace{\int_{-\infty}^\infty y^{\alpha
+\beta}\;e^{-\frac{\hbar^2}{2}ay^2-iyp}\;dy}_{\chi (p)},
\end{eqnarray}
where the integral $\chi(p)$ can be adopted to the integral
\cite{ref:Lebedev}
\begin{eqnarray}
\int_{-\infty}^\infty t^n\;e^{-t^2+2itx}dt=
\frac{\sqrt{\pi}}{(-i)^n 2^n}\;e^{-x^2}H_n(x).
\end{eqnarray}
Thus $\chi (p)$ is straightforward;
\begin{eqnarray}
\chi (p)=(-i)^k\sqrt{\pi}\frac{1}{2^k}\left(
\frac{\sqrt{2}}{\hbar\sqrt{a}}\right)^{k+1}\;
e^{-p^2/{2a\hbar^2}}H_k(p/{\hbar\sqrt{2a}}),
\end{eqnarray}
where $k=\alpha + \beta$. If we use the Rodrigues representation
of the Hermite polynomials
\begin{eqnarray}
H_k(u)=(-1)^k\;e^{u^2}\partial_u^k \;e^{-u^2},
\end{eqnarray}
we get $\chi (p)$ as
\begin{eqnarray}
\chi (p)=\frac{\sqrt{2\pi/a}}{\hbar}(i\partial_p)^{\alpha
+\beta}\;e^{-p^2/{2a\hbar^2}}.
\end{eqnarray}
Thus the WF yields
\begin{eqnarray}\label{sum}
W(q,p)&=&\frac{1}{\hbar\sqrt{2\pi a}}\;e^{-2aq^2}
\sum_{n,m=0}^{\infty}C_n^*\;C_m\;
\sum_{r=0}^{[n/2]}\frac{(-1)^r\,n!}{r!\,(n-2r)!}2^{n-2r}
\sum_{s=0}^{[m/2]}\frac{(-1)^s\,m!}{s!\,(m-2s)!}2^{m-2s}\nonumber\\
& \times & %
\sum_{\alpha=0}^{n-2r}\left(
\begin{array}{c} n-2r \\
\alpha
\end{array}
\right)q^{n-2r-\alpha}(-\frac{\hbar}{2})^\alpha
(i\partial_p)^\alpha \nonumber\\%
& \times & %
\sum_{\beta=0}^{m-2s}\left(
\begin{array}{c} m-2s \\
\beta
\end{array}
\right)q^{m-2s-\beta}(\frac{\hbar}{2})^\beta %
(i\partial_p)^\beta \,e^{-p^2/2a\hbar^2}.
\end{eqnarray}
The last two sums in (\ref{sum}) stand for
$(q-\frac{i\hbar}{2}\partial_p)^{n-2r}$ and
$(q+\frac{i\hbar}{2}\partial_p)^{m-2s}$ respectively. With this
arrangement it is obtained that
\begin{eqnarray}
W(q,p)&=&\frac{1}{\hbar\sqrt{2\pi a}}\;e^{-2aq^2}
\sum_{n=0}^{\infty}C_n^*\sum_{r=0}^{[n/2]}%
\frac{(-1)^r\,n!}{r!\,(n-2r)!}[2(q-\frac{i\hbar}{2}\partial_p)]^{n-2r}\nonumber\\
&\times &%
\sum_{m=0}^{\infty}C_m\sum_{s=0}^{[m/2]}%
\frac{(-1)^s\,m!}{s!\,(m-2s)!}[2(q+\frac{i\hbar}{2}\partial_p)]^{m-2s}\;%
e^{-p^2/{2a\hbar^2}}.
\end{eqnarray}
Thus, with the help of (\ref{power}) and (\ref{hermite}), the
Wigner function can finally be compacted as in the equation
(\ref{main}).

\section{Applications}
\subsection{Gaussian wave packet}
The wave function for a system represented initially by a Gaussian
wave packet is given as
\begin{eqnarray}
\psi(q)=C\,exp\left[-\frac{(q-q_0)^2}{4(\triangle
q)^2}\right]e^{i\,p_0\,q/\hbar},
\end{eqnarray}
where $C=1/[2\pi(\triangle q)^2]^{1/4}$, $\triangle q$ is the
width of the packet centered at $q_0$ and $p_0$ is its average
momentum. According to (\ref{choice})
\begin{eqnarray}
\phi(q)=C\,exp\left[-\frac{q_0^2}{4(\triangle
q)^2}\right]e^{q_0\,q/[2(\triangle q)^2]}\,e^{i\,p_0\,q/\hbar}
\end{eqnarray}
and $a=1/[4(\triangle q)^2]$. Thus
\begin{eqnarray}
\phi^{*}\left(q-\frac{i\hbar}{2}\partial_p\right)
\;\phi\left(q+\frac{i\hbar}{2}\partial_p\right)=C^2\,exp\left[-\frac{q_0^2}{2(\triangle
q)^2}\right]e^{q_0\,q/(\triangle q)^2}\,e^{-p_0\,\partial_p},
\end{eqnarray}
which is obviously a translation operator which converts $p$ to
$p-p_0$. Therefore (\ref{main}) and the fact that $\triangle
q\triangle p=\hbar/2$ yield the WF as
\begin{eqnarray}
W(q,p)=\frac{1}{\pi\hbar}\,exp\left[-\frac{(q-q_0)^2}{2(\triangle
q)^2}\right]\,exp\left[-\frac{(p-p_0)^2}{2(\triangle p)^2}\right]
\end{eqnarray}
that confirms the correct result.
\subsection{Harmonic oscillator}
The wave function corresponding to the Hamiltonian
$\hat{H}=\hat{p}^2/2+\hat{q}^2/2$ is given by
\begin{eqnarray}
\psi_n(q)=C_n\;e^{-q^2/2\hbar}\;H_n(q/\sqrt{\hbar}),
\end{eqnarray}
where $C_n=[1/(\pi\hbar)]^{1/4}/\sqrt{2^n n!}\,$. By virtue of the
equation \cite{ref:Tegmen}
\begin{eqnarray}
&&H_n\left[\frac{1}{\sqrt{\hbar}}\left(q-\frac{i\hbar}{2}\partial_p\right)\right]
H_n\left[\frac{1}{\sqrt{\hbar}}\left(q+\frac{i\hbar}{2}\partial_p\right)\right]
e^{-p^2/\hbar}\nonumber \\
&&=(-1)^n\; 2^n\; n!\; L_n [(2\,q^2+2\,p^2)/\hbar]\;e^{-p^2/\hbar}
\end{eqnarray}
which is obtained by iteration, (\ref{main}) gives the well known
WF
\begin{eqnarray}
W_n(q,p)=\frac{(-1)^n}{\pi\hbar}\,e^{-2H/\hbar}L_n(4H/\hbar),
\end{eqnarray}
where $H=p^2/2+q^2/2$ and $L_n$ denotes the Laguerre polynomial of
order $n$.
\subsection{Singular oscillator}
The system is described by the Hamiltonian
$\hat{H}=\hat{p}^2/2+\hat{q}^2/2+g^2/\hat{q}^2$ with $g$ is a real
constant. The normalized eigenfunctions of the Hamiltonian are
\begin{eqnarray}
\psi_n(q)=C_n\,q^\alpha\,L_n^{\alpha-1/2}(q^2/\hbar)\,e^{-q^2/{2\hbar}},
\end{eqnarray}
where $C_n^2=n!/[\Gamma(n+\alpha+1/2)\,\hbar^{\alpha+1}]$,
$\Gamma$ is the Gamma function, $L_n^{\alpha+1/2}$ is the
associated Laguerre polynomial and $\alpha =1/2+(1/4+2g^2)^{1/2}$
\cite{ref:Perelomov}. The operator equation version of the WF is
straightforward with the help of the argument presented up to now;
\begin{eqnarray}
W_n(q,p)&=&C_n^2\,e^{-q^2/\hbar}\left(
q^2+\frac{\partial_p^2}{4}\right)^\alpha\nonumber \\%
&\times&L_n^{\alpha-1/2}\left[\frac{1}{\hbar}\left(q-\frac{i\partial_p}{2}\right)^2\right]
L_n^{\alpha-1/2}\left[\frac{1}{\hbar}\left(q+\frac{i\partial_p}{2}\right)^2\right]
e^{-p^2/\hbar}.
\end{eqnarray}
For an arbitrary $\alpha$ and $n$, integral tables and the
approach discussed here fail to get an exact WF of this system.
Especially for the fractional values of $\alpha$, even determining
the ground state carries big difficulty since the action of the
operator $(q^2+\partial_p^2/4)^\alpha$ on $exp(-p^2/\hbar)$ is
unknown. (At least, I am not aware of it). But for some special
values of $\alpha$ and $n$, WFs for this system can be obtained
explicitly \cite{ref:Tegmen}.

As a limit of the applicability of the method, consider the one
dimensional anyon system \cite{ref:Antonyan}, where the wave
function
\begin{eqnarray}
\psi_n(q)\quad\propto\quad e^{-q^2/2}\;q^{1/2}\;H_n(q)
\end{eqnarray}
corresponds to the Hamiltonian
$\hat{H}=\hat{p}^2/2-1/\hat{q}-1/\hat{q}^2$. Obviously the matter
arises from the term $q^{1/2}$ since the smoothness condition of
$\phi (q)$ is violated.\\\\
{\bf Acknowledgments} \\\\
The author wishes to express his
appreciations to T. Altanhan and B. S. Kandemir for their helpful
discussions. This work was supported in part by the Scientific and
Technical Research Council of Turkey (T\"{U}B\.{I}TAK).


\begin{thebibliography}{0}
\bibitem{ref:Wigner} {Wigner~E~P.},
   {\it Phys. Rev}. 40 (1932) 74.
\bibitem{ref:Hillery} {Hillery~M., O'Connel~R~F.,
   Scully~M~O. and Wigner~E~P.},
  {\it Phys. Rep}. 106 (1984) 121.
\bibitem{ref:Lebedev} {Lebedev~N~N.},
  {\it Special functions and their applications},
   (Dover Publications) (1972).
\bibitem{ref:Tegmen} {Tegmen~A., Altanhan~T.
   and Kandemir~B~S.}, {\it Eur Phys J D} 41 (2007) 397-402.
\bibitem{ref:Perelomov} {Perelomov~A.},
  {\it Generalized coherent states and their applications},
   (Springer, Berlin) (1986).
\bibitem{ref:Antonyan} {Hakobyan~Y. and Ter-Antonyan~V.},
  {\it Phys. Atom. Nucl+}. 68 (2005) 1709.
\end{thebibliography}
\end{document}